\documentclass{LMCS} 

\usepackage{enumerate}
\usepackage{hyperref}

\usepackage{eucal,,amssymb, amsbsy,amsthm}

\newcommand\cX{\ol{X}}
\newcommand\cx{\ol{x}}
\newcommand\ol[1]{\overline{#1}}
\newcommand\euro{\kern1pt\text{{\sffamily c}\kern-5.5pt\rule[2.7pt]{4.5pt}{.4pt}\kern-4.5pt\rule[1.8pt]{4.3pt}{.4pt}\kern1pt}}
\newcommand\Neut{\EuScript{N}}
\newcommand\seq{\subseteq}

\newcommand{\bin}{{\rm bin}}
\newcommand{\mi}{{\rm min}}
\newcommand{\ma}{{\rm max}}

\newcommand\AC{{\rm AC}^0}
\newcommand\ATIME{{\rm ATIME}}
\newcommand{\pspace}{{\rm PSPACE}}
\newcommand{\np}{{\rm NP}}
\newcommand{\tc}{{\rm TC}^0}

\newcommand{\SAC}{{\rm SAC^1}}
\newcommand{\NC}{{\rm NC^1}}
\newcommand\dspace{\text{\rm DSPACE}}
\newcommand\dtime{\text{\rm DTIME}}

\newcommand{\tally}{{\rm tally}}
\newcommand\REG{{\rm REG}}
\newcommand\LOGCFL{{\rm LOGCFL}}
\newcommand\CFL{{\rm CFL}}
\newcommand\DLOG{{\rm DLOGTIME}}
\newcommand{\ASPACETS}{{\rm ASPACE\text{-}TREESIZE}}
\newcommand{\NSPACE}{{\rm NSPACE}}
\newcommand\ACC{{\rm ACC}^0}
\newcommand\modPH{{\rm Mod\text{-}PH}}
\newcommand\lmodPH{{\rm Mod\text{-}LinH}}
\newcommand\CH{{\rm CH}}
\newcommand\lCH{{\rm Lin\text{-}CH}}
\newcommand\Modq{{\rm Mod_qP}}
\newcommand\ph{{\rm PH}}
\newcommand\pp{{\rm PP}}

\newcommand{\bit}{{\rm BIT}}
\newcommand\CA{{\EuScript{A}}}
 
\newcommand{\NN}{\mathbb{N}}
\newcommand{\dom}{{\rm dom}}
\newcommand{\MOD}{{\rm MOD}_q}
\newcommand{\Maj}{{\rm Maj}}

\newcommand\FO{\text{\rm FO}}

\newcommand\SOM{{\rm SOM}}

\newcommand\Mon{\text{\rm Mon}}
\newcommand\Grp{\text{\rm Grp}}
\newcommand\QMon{Q_{\Mon}}
\newcommand\QGrp{Q_{\Grp}}
\newcommand\msQ{\text{\rm mon-}Q^1}

\newcommand\msQMon{\msQ_{\Mon}}
\newcommand\msQGrp{\msQ_{\Grp}}
\newcommand\QmodM{\text{\rm mon-}Q^{\star}_{\Mon}}
\newcommand\Qmod{\text{\rm mon-}Q^{\star}_{\Grp}}
\newcommand\QmodL{\text{\rm mon-}Q^{\star}_L}
\newcommand\QmodH{\text{\rm mon-}Q^{\star}_{{\rm pad}(H)}}
\newcommand\msQH{\text{\rm mon-}Q^1_{{\rm pad}(H)}}

\newcommand\qfree{\text{\rm QF}}
\newcommand\QmodMaj{\text{\rm mon-}Q^{\star}_{\rm Maj}}

\newcommand\msQC{\text{\rm mon-}Q^1_{\mathcal{C}}}
\newcommand\QmodC{\text{\rm mon-}Q^{\star}_{\mathcal{C}}}
\newcommand\QmodMS{\text{\rm mon-}Q^{\star}_{\REG^s}}
\newcommand\QmodS{\text{\rm mon-}Q^{\star}_{\CFL^s}}

\newcommand\Leaf{{\rm Leaf}}

\newcommand\LeafP{\Leaf^{\rm P}}

\newcommand\leafM{{\rm leafstring}^{M}}

\newcommand\LeafFA{\Leaf^{\text{\rm FA}}}

\def\doi{6 (3:25) 2010}
\lmcsheading%
{\doi}
{1--22}
{}
{}
{Sep.~21, 2009}
{Sep.~20, 2010}
{}

\begin{document}

\title[On Second-Order Monadic Monoidal and Groupoidal Quantifiers]{On Second-Order Monadic Monoidal and Groupoidal Quantifiers\rsuper*}

\author[J.~Kontinen]{Juha Kontinen\rsuper a}
\address{{\lsuper a}Department of Mathematics and Statistics, University of Helsinki, P.O. Box 68, FI-00014 University of Helsinki, Finland}
\email{juha.kontinen@helsinki.fi}
\thanks{{\lsuper a}Supported by grant 127661 of the Academy of Finland}

\author[H.~Vollmer]{Heribert Vollmer\rsuper b}
\address{{\lsuper b}Institut f\"ur Theoretische Informatik, Universit\"at
Hannover, Appelstra\ss e 4, 30167 Hannover, Germany}
\email{vollmer@thi.uni-hannover.de}
\thanks{{\lsuper b}Supported partially by DFG grants VO 630/6-1 and 6-2}

\titlecomment{{\lsuper*}A previous version of this paper appeared in the Proceedings of the Workshop on Logic, Language, Information and Computation 2008, Springer Lecture Notes in Computer Science Vol.~5110, pp.~238--248, Springer Verlag, 2008.}

\begin{abstract}
We study logics defined in terms of second-order monadic monoidal and groupoidal quantifiers.
These are generalized quantifiers defined by monoid and groupoid
word-problems, equivalently, by regular and context-free languages. We give a computational classification 
of the expressive power of these logics over strings with varying built-in predicates. In particular, we show that  
$\ATIME(n)$ can be logically characterized in terms of  second-order monadic monoidal quantifiers. 
\end{abstract}

\maketitle

\section{Introduction}

\noindent We study logics defined in terms of so-called second-order
monadic monoidal and groupoidal quantifiers.  These are generalized
quantifiers defined by monoid and groupoid word-problems,
equivalently, by regular and context-free languages.  A
\emph{groupoid} is a finite multiplication table with an identity
element.  For a fixed groupoid $G$, each $S\subseteq G$ defines a
$G$-word-problem, i.e., a language $\mathcal{W}(S,G)$ composed of all
words $w$, over the alphabet $G$, that can be bracketed in such a way
that $w$ multiplies out to an element of $S$. The word-problem of a
\emph{monoid}, i.e., an associative groupoid, is defined analogously.
Groupoid word-problems relate to context-free languages in the same
way as monoid word-problems relate to regular languages: every such
word-problem is context-free, and every context-free language is a
homomorphic pre-image of a groupoid word-problem (this result is
credited to Valiant in \cite{belemc93}).

In descriptive complexity, (first-order) monoidal quantifiers have been studied extensively in 
connection to the complexity class $\NC$ and its sub-classes (see  \cite{baimst90,bacostth92,str94, stthth95}). However, in
 order to define  non-regular languages in terms of monoidal quantifiers, some built-in relations, in addition to $<$, need to be assumed. It was shown already in \cite{baimst90} that first-order logic with unnested unary  monoidal quantifiers characterizes the class of regular languages, $\REG$,  over strings without auxiliary built-in relations. This characterization of $\REG$ was generalized in  \cite{lamcscvo01} to allow also non-unary monoidal quantifiers, even  with arbitrary nestings.  In \cite{MR1908783},  the same was shown to hold   for 
 second-order monadic monoidal  quantifiers:
\begin{equation}\label{monoidcase}
 \msQMon\FO \equiv \SOM(\msQMon) \equiv \REG\equiv \exists \SOM.
\end{equation}
In \eqref{monoidcase}, $\exists\SOM$ denotes  existential second-order monadic logic and the  logic $\msQMon\FO$ consists of all formulas in which 
a monadic second-order monoidal quantifier $Q^1_L$ is applied to 
an appropriate tuple of $\FO$-formulas without further occurrences of
second-order quantifiers. On the other hand, in $\SOM(\msQMon)$
 arbitrary nestings of monoidal quantifiers are allowed. Here a crucial assumption is that there are no auxiliary built-in relations, besides the order, since already $\SOM(+)$, i.e.,  second-order monadic logic with built-in  addition, defines exactly the languages in the linear fragment of the polynomial hierarchy \cite{MR1465635}. 

We see that with monoidal quantifiers the situation is clear-cut, i.e., formulas with
monadic second-order monoidal  quantifiers cannot define non-regular
languages. On the other hand,  over strings with built-in arithmetic  (i.e.,  built-in $+$ and $\times$) the classes in \eqref{monoidcase} are presumably not
equal, e.g., $\exists \SOM\subseteq \np$  and already in $\msQMon\FO(+,\times)$ $\pspace$-complete languages can be defined as we show below in     Corollary \ref{monpspace}.

In \cite{belemc93}, the  elaborate theory connecting  monoids  to the fine structure of  $\NC$ was generalized 
to groupoids and $\LOGCFL$. It was shown  in \cite{belemc93} that there exists a single groupoid whose word-problem is complete for $\LOGCFL$ under $\DLOG$-reductions, implying also a logical characterization for $\LOGCFL$ in terms of first-order groupoidal quantifiers. Building on this result, a systematic investigation of  first-order groupoidal quantifiers was initiated in  \cite{lamcscvo01}. 

In \cite{MR1908783} it was asked what is the relationship of the corresponding (second-order) logics  if
 monoidal quantifiers are replaced by groupoidal quantifiers in \eqref{monoidcase}.  Here we  address this question and show the following (see Corollary \ref{cor}):
\begin{equation}\label{main}
 \msQGrp\FO(+,\times)\equiv \SOM(\msQGrp) .
\end{equation}
It is interesting to note that for  groupoidal quantifiers we have a similar collapse result as for monoidal quantifiers, but this time assuming  built-in arithmetic on the left. Note that, over ordered structures, the relations $+$ and $\times$ are definable in the logic $\SOM(\msQGrp)$  (see \cite{baimst90} and \cite{lamcscvo01}).   It is  an open question whether the built-in relations $+$ and $\times$ are really needed  for the equivalence in  \eqref{main} to hold.

In the literature,   second-order monadic quantifiers have been studied under two slightly different  semantics
(for each $L$, quantifiers $Q^1_L$ and   $Q^{\star}_L$).  We will  show
that the analogue of \eqref{main} for the alternative  semantics $Q^{\star}_L$ remains valid even if we drop the  built-in predicates $+$ and $\times$ from $\Qmod\FO(+,\times)$, i.e.,
\begin{equation}\label{mainr2}
 \Qmod\FO\equiv \SOM(\Qmod) .
\end{equation}
Since the logics in  \eqref{main} and \eqref{mainr2} are all equivalent (see Corollary \ref{cor}), it follows that the 
only remaining open question regarding the equivalences between  logics with groupoidal quantifiers  is whether
\[\msQGrp\FO \equiv \Qmod\FO?\]
This question is directly concerned with the problem of pinning down the exact expressive power of the so-called finite leaf automata with
context-free leaf languages (see Theorem \ref{neutral} and Corollary \ref{Grppspace}). 

In this paper we aim for a concise classification of the expressive power of the  logics with second-order monadic monoidal and groupoidal quantifiers.
We first note that the difference  between the two semantics, i.e., $Q^1_L$ and $Q^{\star}_L$, disappears assuming  built-in arithmetic. This already simplifies the picture considerably. However, especially in the monoidal case, the expressive power of the quantifiers  $Q^{\star}_L$ without built-in arithmetic remains open.  For groupoidal quantifiers, we find that 
\[ \Qmod\FO\equiv \SOM(\Qmod) \equiv 2^{\LOGCFL},\]
where  $2^{\LOGCFL}$ equals the class of languages whose tally version resides in $\LOGCFL$.
For monoidal quantifiers, we show that  
\[\SOM(\QmodM,+,\times) \equiv \ATIME(n).\]
Table \ref{table:summary} below contains a  summary of  our complexity results. 

\section{Preliminaries}

\noindent We follow standard notation for second-order monadic logic
with linear order, see, e.g., \cite{str94}. We mainly restrict our
attention to {\em string structures}, i.e., structures of {\em string
  signatures} $\tau=\langle P_{a_1},\dots,P_{a_s}\rangle$, where all
the predicates $P_{a_i}$ are unary. We assume that the universe
$\dom(\CA)$ of each structure $\CA$ is of the form $\{0,\dots,n-1\}$
and that the logic's linear order symbol refers to the numerical order
on $\{0,\dots,n-1\}$. We restrict attention to structures $\CA$ in
which the interpretations $P^{\CA}_{a_i}$ of the predicates $P_{a_i}$
satisfy the following: $P^{\CA}_{a_i}\cap P^{\CA}_{a_j}=\emptyset$,
for $i \neq j$, and $\cup_{1\le i\le s} P^{\CA}_{a_i}=\dom(\CA)$. Such
$\tau$-structures correspond to strings over the alphabet
$\{a_1,\dots,a_s\}$ in the usual way.

An alphabet $\Sigma$ is a finite set of symbols. For technical reasons to be motivated shortly, we assume that every alphabet  has a built-in linear
order, and, to indicate that order, we write alphabets as sequences of symbols, e.g., in the above case we write $(a_1,\dots,a_s)$. 
The set of all finite $\Sigma$-strings is denoted by $\Sigma^*$
and $\Sigma^+=\Sigma^*\setminus \{\epsilon\}$, where $\epsilon$ is the empty string. For a string $w$, $|w|$ denotes the length of $w$ and   $|w|_a$  the number of occurrences of  the letter $a$ in $w$. The concatenation of the strings $w$ and $w'$ is denoted by $w^{\smallfrown} w'$, and $a^k$ denotes the string $b_1\cdots b_k$, where $b_i=a$ for $1\le i\le k$.  
For $L\subseteq \Sigma^*$ and $e\in \Sigma$, the letter $e$ is a {\em neutral letter} of $L$ if  for all $u,v\in\Sigma^*$, we have $uv\in L
\iff ue v\in L$. The class of languages that have a neutral letter is denoted by $\Neut$.

For a signature $\tau =\langle P_{a_1},\dots,P_{a_s}\rangle$, the first-order $\tau$-formulas, $\FO[\tau]$, are built
 from first-order variables in the usual way, using the Boolean connectives $\{ \wedge, \vee, \neg
\}$, the  predicates $P_{a_i}$ together with $\{ = , <\}$, the
constants $\mi$ and $\ma$, the first-order  quantifiers
$\{\exists, \forall \}$, and parentheses. $\SOM[\tau]$ extends $\FO[\tau]$ in terms of
unary second-order variables and second-order quantifiers $\{\exists, \forall \}$.
(The letters SOM stand for second order monadic logic; in
the literature, this logic is sometimes denoted by MSO.) 

For a complexity class $\mathcal{C}$ and logics $\mathcal{L}$ and $\mathcal{L}'$, we write 
 $ \mathcal{L}\le\mathcal{L}'$ 
 if for every string signature $\tau$ (unless otherwise specified), and every sentence $\varphi\in \mathcal{L}[\tau]$ there is an equivalent sentence  $\psi\in \mathcal{L}'[\tau]$. Analogously, we write  $\mathcal{L}\le \mathcal{C}$ if the class of languages, over any alphabet, which can be defined in $\mathcal{L}$ is contained in  $\mathcal{C}$. We write $ \mathcal{L}\equiv\mathcal{L}'$   ($\mathcal{L}\equiv \mathcal{C}$)  if  $\mathcal{L}\le\mathcal{L}'$  and  $ \mathcal{L}'\le\mathcal{L}$ ( $\mathcal{L}\le\mathcal{C}$  and  $ \mathcal{C}\le\mathcal{L}$).
It is known \cite{mcpa71} that $\FO$ is equal to the class
of star-free regular languages and that $\SOM\equiv\REG$, where $\REG$  is the class of  regular languages (see \cite{buel58,bue62,tra61}).

Sometimes we assume that our structures (and logics) are  equipped with auxiliary built-in
predicates in addition to $<$, e.g.,  the ternary predicates $+$ and $\times$. The predicates  $+$ and $\times$ are defined as
\begin{eqnarray*}
+(i,j,k) &\Leftrightarrow& i+j=k,\\
\times(i,j,k) &\Leftrightarrow& i\times j=k. 
\end{eqnarray*}
The predicate $\bit$ is a further important predicate which is defined  by: $\bit(a,j)$ holds iff the bit with weight  $2^{j}$ is $1$ in the binary representation of $a$. The presence of  built-in predicates is signalled, e.g., by  the notation $\FO(+,\times)$ and $\FO(Q_L, +,\times)$. It is well known that  $\FO(+,\times)\equiv \FO(\bit)$ (see \cite{imm99}).   In fact,
it was shown in \cite{DDLW} that $\bit$ alone can define the corresponding canonical ordering, i.e.,  the symbol $<$  can  dropped from  $\FO(\bit)$ without a loss in expressive power.

\subsection{Generalized quantifiers}
Next, we extend logics in terms of generalized quantifiers. The Lindstr{\"o}m quantifiers of Definition \ref{grquant}
are precisely what has been referred to as ``Lindstr{\"o}m quantifiers on strings''
\cite{buvo98}. The original more general definition \cite{lin66} uses
transformations to arbitrary structures, not necessarily of string
signature.
\begin{defi}\label{grquant}
Consider a language $L$ over an alphabet $\Sigma = (a_1, a_2, \dots,
a_{s})$. Such a language gives rise to a Lindstr{\"o}m quantifier $Q_L$,
that may be applied to any sequence of $s-1$ formulas as follows:

Let $\ol{x}$ be a $k$-tuple of pairwise distinct variables. Let $\CA$  be a structure and $\dom(\CA)=\{ 0,1,\dots,n-1\}$. 
We assume the
lexicographic ordering on $\{0,1,\dots,n-1\}^k$, and we write $\cx^{(0)} <
\cx^{(1)} < \dots < \cx^{(n^k-1)}$ for the sequence of potential values
taken on by $\ol{x}$.  The $k$-ary \emph{Lindstr{\"o}m quantifier} $Q_L$
binding $\ol{x}$ takes a meaning if $s-1$ formulas, each having as
free variables the variables in $\ol{x}$ (and possibly others), are
available.  Let $\varphi_1(\ol{x})$, $\varphi_2(\ol{x})$, $\dots$,
$\varphi_{s-1}(\ol{x})$ be these $s-1$ formulas.  Then
$$\CA\models Q_L\ol{x}\bigl[ \varphi_1(\ol{x}), \varphi_2(\ol{x}), \dots,
\varphi_{s-1}(\ol{x})\bigr]$$
 iff the word of length $n^k$ whose $i$th letter, $0\leq i\leq n^k-1$, is
\begin{displaymath}
\left\{
   \begin{array}{ll}
       a_1 & \mbox{if $\CA \models \varphi_1(\cx^{(i)})$,} \\
       a_2 & \mbox{if $\CA \models \neg\varphi_1(\cx^{(i)})
				\wedge\varphi_2(\cx^{(i)})$,} \\
        &\vdots\\
       a_{s}& \mbox{if  $\CA \models \neg\varphi_1(\cx^{(i)})
			\wedge \neg\varphi_2(\cx^{(i)}) \wedge\dots
			\wedge \neg\varphi_{s-1}(\cx^{(i)})$,}
   \\
   \end{array}
       \right.
\end{displaymath}
belongs to $L$.
\end{defi}

As an example, take $s=2$ and consider $L_{\exists} :=
0^*1(0+1)^*$; then $Q_{L_\exists}$ is the usual first-order
existential quantifier. Similarly, the universal quantifier can be
expressed using the language $L_{\forall}:= 1^*$. Finally,  for  $p>1$ and  $L_{\text{mod }p}=\{w\in \{0,1\}^*\ |\ |w|_1\equiv 0\ (\text{mod }p) \}$,  the quantifiers $Q_{L_{\text{mod }p}}$ are known as modular counting
quantifiers \cite{str94}.

\begin{defi}Let $\tau$ be a signature,  $L$  a language over an alphabet $\Sigma = (a_1, a_2, \dots,a_{s})$, and $\mathcal{C}$ a class of languages.
\begin{enumerate}[$\bullet$]
\item The set of $\tau$-formulas, $Q_L\FO[\tau]$, of the logic $Q_L\FO$ consists of all formulas of the form
 $$Q_L\ol{x}\bigl[ \varphi_1(\ol{x}), \varphi_2(\ol{x}), \dots,
\varphi_{s-1}(\ol{x})\bigr],$$
where, for some $k$, $\ol{x}$ is a $k$-tuple of pairwise distinct variables, and  $\varphi_i(\ol{x})$ is a $\FO[\tau]$-formula for $1\le i \le s-1$.

\item The set of $\tau$-formulas,  $\FO(Q_L)[\tau]$,  of the logic $\FO(Q_L)$  is defined by extending the formula formation rules of  $\FO$ by the following clause:
if, for some $k$, $\ol{x}$ is a $k$-tuple of pairwise distinct  variables, and $\varphi_i(\ol{x})$ is a formula for $1\le i \le s-1$, then 
 $$Q_L\ol{x}\bigl[ \varphi_1(\ol{x}), \varphi_2(\ol{x}), \dots,
\varphi_{s-1}(\ol{x})\bigr]$$
is a formula, too.
\item Define the sets of $\tau$-formulas of the logics $Q_{\mathcal{C}} \FO$ and $\FO(Q_{\mathcal{C}})$ by
\begin{eqnarray*}\label{logics}
 Q_{\mathcal{C}} \FO[\tau]&:=& \bigcup_{L\in \mathcal{C}} Q_L \FO[\tau],\\
\FO(Q_{\mathcal{C}})[\tau]&:=& \bigcup_{L\in \mathcal{C}}\FO(Q_L)[\tau].
\end{eqnarray*}
\end{enumerate}
\end{defi}
\noindent
In this article we are especially interested in quantifiers defined by monoid and groupoid word-problems.
\begin{defi}\label{grpmon}
A {\em groupoidal quantifier\/} ({a \em monoidal quantifier\/}) is a Lindstr{\"o}m quantifier $Q_L$
where $L$ is a word-problem of some finite groupoid (monoid).
The usage of groupoidal quantifiers and monoidal quantifiers in our logical language is signalled by the subscripts $\Grp$ and $\Mon$, respectively. We define 
\begin{eqnarray*}
\QGrp\FO:=Q_{\mathcal{C}}\FO &\mbox{ }& \FO(\QGrp):= \FO(Q_{\mathcal{C}})\\ 
\QMon\FO:=Q_{\mathcal{C'}}\FO &\mbox{ }& \FO(\QMon):= \FO(Q_{\mathcal{C'}}),
\end{eqnarray*}
where $\mathcal{C}$ ($\mathcal{C}'$) is the class of all  word-problems of finite groupoids (monoids).   
\end{defi}

 Second-order Lindstr{\"o}m quantifiers on strings were introduced in
\cite{buvo98}. Here, we are mainly interested in those binding only
set variables (i.e., unary relations), so-called \emph{monadic quantifiers}. For each language $L$,
we define two monadic quantifiers $Q^1_L$ and $Q^{\star}_L$ with slightly different
interpretations. It turns out that the interpretation $Q^1_L$, which was used in \cite{MR1908783},
is natural in the context of finite leaf automata. On the other hand, the quantifier
$Q^{\star}_L$ is the exact second-order analogue of the corresponding first-order quantifier $Q_L$.

\begin{defi}\label{lind2}
Consider a language $L$ over an alphabet $\Sigma = (a_1, a_2, \dots,
a_{s})$. Let $\ol{X} = (X_1,\dots,X_k)$ be a $k$-tuple of   pairwise distinct  unary
second-order variables and let $\CA$  be a structure with $\dom(\CA)=\{0,1,\dots,n-1\}$.   There are $2^{nk}$
different instances (assignments) of $\ol{X}$ over $\CA$. We assume the following
ordering on those instances: Let each instance of a single $X_i$ be
encoded by the bit string $s^i_0\cdots s^i_{n-1}$ with the meaning $s^i_j =
1 \iff j\in X_i$. Then
\begin{enumerate}[(1)]
\item\label{a} we encode an instance of $\ol{X}$ by the bit
string
\[s^1_0s^2_0\cdots s^k_0s^1_1s^2_1\cdots s^k_1\cdots s^1_{n-1}s^2_{n-1}\cdots s^k_{n-1}\]
and order the instances lexicographically by
their codes.
\item\label{b} we  encode an instance of $\ol{X}$ by the bit
string
\[s^1_0s^1_1\cdots s^1_{n-1}s^2_0s^2_1\cdots s^2_{n-1}\cdots s^k_0s^k_1\cdots s^k_{n-1}\]
 and order the instances lexicographically by
their codes.
\end{enumerate}
The \emph{monadic second-order Lindstr{\"o}m quantifier\/} $Q^1_L$ (respectively $Q^{\star}_L$) binding $\ol{X}$ takes
a meaning if $s-1$ formulas, each having free variables $\ol{X}$, are
available.  Let $\varphi_1(\ol{X})$, $\varphi_2(\ol{X})$, $\dots$,
$\varphi_{s-1}(\ol{X})$ be these $s-1$ formulas.  Then 
$$\CA\models Q^1 _L\ol{X}\bigl[ \varphi_1(\ol{X}), \varphi_2(\ol{X}), \dots,
\varphi_{s-1}(\ol{X})\bigr]$$
 iff the
word of length $2^{nk}$ whose $i$th letter, $0\leq i\leq 2^{nk}-1$, is
\begin{displaymath}
 \left\{
   \begin{array}{ll}
       a_1  & \mbox{if $\CA \models \varphi_1(\cX^{(i)})$,} \\
       a_2  & \mbox{if $\CA \models
                    \neg\varphi_1(\cX^{(i)})\wedge\varphi_2(\cX^{(i)})$,} \\
            & \vdots\\
       a_{s}& \mbox{if  $\CA \models \neg\varphi_1(\cX^{(i)})
		\wedge\neg\varphi_2(\cX^{(i)}) \wedge\dots
		\wedge\neg\varphi_{s-1}(\cX^{(i)})$,} \\
   \end{array}
 \right.
\end{displaymath}
belongs to $L$. Above, $\cX^{(0)} < \cX^{(2)} < \dots < \cX^{(2^{nk}-1)}$ denotes the sequence
of all instances ordered as in \eqref{a}. The notation $Q^{\star}_L$ is used when the instances are ordered according to  \eqref{b}.
\end{defi}
Again, taking as examples the languages $L_{\exists}$ and
$L_{\forall}$, we obtain the usual second-order monadic existential and
universal quantifiers. Note that for $L\in \{L_{\exists}, L_{\forall}\}$ the quantifiers $Q^1 _L$
and $Q^{\star}_L$ are ``equivalent''. This is due to the fact that,  for the
membership in $L$, the order of the letters in a word does not matter.

\begin{defi}Let $\tau$ be a signature,  $L$  a language over an alphabet $\Sigma = (a_1, a_2, \dots,a_{s})$, and $\mathcal{C}$ a class of languages.
\begin{enumerate}[$\bullet$]
\item The set of $\tau$-formulas, $\msQ_L\FO[\tau]$, of the logic  $\msQ_L\FO$  consists of 
all formulas of the form
\begin{equation}\label{SOlogics}
Q^1 _L\ol{X}\bigl[ \varphi_1(\ol{X}), \varphi_2(\ol{X}), \dots,
\varphi_{s-1}(\ol{X})\bigr],              
\end{equation}  
where, for some $k$, $\ol{X}$ is a $k$-tuple of pairwise distinct  unary second-order variables, and  $\varphi_i(\ol{X})$ is a $\FO[\tau]$-formula with variables $\ol{X}$, for $1\le i\le s-1$. 

\item The set of $\tau$-formulas, $\SOM(\msQ_L)[\tau]$,  of the logic $\SOM(\msQ_L)$ is defined by extending the formula formation rules of $\SOM[\tau]$ by the following clause:
if, for some $k$, $\ol{X}$ is a $k$-tuple of pairwise distinct  unary second-order variables, and $\varphi_i(\ol{X})$ is a formula for $1\le i\le s-1$, then 
\[Q^1 _L\ol{X}\bigl[ \varphi_1(\ol{X}), \varphi_2(\ol{X}), \dots,
\varphi_{s-1}(\ol{X})\bigr] \]     
is a formula, too.

\item Define the sets of $\tau$-formulas of the logics $\msQC \FO$ and $\SOM(\msQC)[\tau]$ by
\begin{eqnarray*}
 \msQC \FO[\tau]&:=& \bigcup_{L\in \mathcal{C}} \msQ_L\FO[\tau],\\
\SOM(\msQC)[\tau]&:=& \bigcup_{L\in \mathcal{C}}\SOM(\msQ_L)[\tau].\\
\end{eqnarray*}

\item The logics $\QmodL\FO$, $\SOM(\QmodL)$, $\QmodC \FO$, and $\SOM(\QmodC)$  are defined analogously by replacing
$Q^1 _L$ everywhere with $Q^{\star}_L$.
\end{enumerate}
\end{defi}
\noindent
Analogously to the first-order case (see Definition \ref{grpmon}), we
use the subscripts  $\Grp$ and $\Mon$ to indicate that all  groupoidal quantifiers or monoidal quantifiers are available in the corresponding logic, e.g., $\SOM(\msQGrp):= \SOM(\msQC)$, where  $\mathcal{C}$  is the class of all  word-problems of finite groupoids.

The next proposition shows that the difference between the two semantics of second-order monadic quantifiers disappears in the presence of built-in arithmetic (or if the  arithmetic predicates are definable). Below, we write $\psi^{\CA}$ for the relation defined by the formula $\psi$ in a structure $\CA$, i.e., if $\psi$ has  $k$ free variables, then     
\[ \psi^{\CA}=\{ \overline{a}\in \dom(\CA)^k\ |\ \CA\models \psi(\overline{a} ) \}.    \] 

\begin{lem}\label{suffle} Let $X_1,\ldots,X_k$ be  unary second-order variables. There are $\FO(+,\times)$-formulas
 $\phi_1(x,\overline{X}), \ldots,\phi_k(x,\overline{X})$ and  $\psi_1(x,\overline{X}), \ldots,\psi_k(x,\overline{X})$ such that for all  $\CA$  and  $A_1,\ldots,A_k\subseteq \dom(\CA)=\{0,1,\ldots, n-1\}$, (where  $A_i$ is encoded by $s^i_0\cdots s^i_{n-1}$ as in Definition \ref{lind2}) it holds that the encoding of  $(\phi_1^{(\CA,\overline{A})}, \ldots,\phi_k^{(\CA,\overline{A})})$ as a bit string as in clause \ref{a} of Definition \ref{lind2} results with 
\begin{equation}\label{fs}
 s^1_0s^1_1\cdots s^1_{n-1}s^2_0s^2_1\cdots s^2_{n-1}\cdots s^k_0s^k_1\cdots s^k_{n-1},
\end{equation}
and the encoding of $(\psi_1^{(\CA,\overline{A})}, \ldots,\psi_k^{(\CA,\overline{A})})$ as a bit string as in clause \ref{b} of Definition \ref{lind2} results with 
\begin{equation}\label{sf}
s^1_0s^2_0\cdots s^k_0s^1_1s^2_1\cdots s^k_1\cdots s^1_{n-1}s^2_{n-1}\cdots s^k_{n-1}.
\end{equation}
\end{lem}
\proof 
Let us show how to construct the formulas  $\phi_1(x,\overline{X}) \ldots,\phi_k(x,\overline{X})$. The idea simply is that 
\[ \CA\models \phi_i(j,\overline{A}) \]
should hold if the bit in position $jk+i$ from the left in \eqref{fs}  is $1$ if and only if  $jk+i=(c-1)n+r$ and $r-1\in A_c$, where
$0\le j\le n-1$, $ 1\le i\le k $, $ 1\le c\le k $, and  $1\le r\le n$. This condition can be easily expressed in $\FO(+,\times)$. The formulas  $\psi_1(x,\overline{X}) \ldots,\psi_k(x,\overline{X})$ can be constructed completely analogously. 

Note also that the formulas  $\phi_1(x,\overline{X}), \ldots,\phi_k(x,\overline{X})$ define a  permutation of $k$-tuples of unary relations  and that  $\psi_1(x,\overline{X}), \ldots,\psi_k(x,\overline{X})$ define the  inverse of this permutation.
\qed

\begin{prop}\label{sufflecor}For any $L$, $\msQ_L\FO(+,\times)\equiv \QmodL\FO(+,\times)$.
\end{prop}
\proof
 by Lemma \ref{suffle},  $Q^{\star}_L\ol{X}\bigl[ \varphi_1(\ol{X}), \dots,
\varphi_{s-1}(\ol{X})\bigr]$ can be expressed as 
\[Q^1 _L\ol{X}\bigl[ \varphi_1(X_1/\psi_{1}(\ol{X}),\ldots, X_k/\psi_{k}(\ol{X}) ), \dots,
\varphi_{s-1}(X_1/\psi_{1}(\ol{X}),\ldots, X_k/\psi_{k}(\ol{X}) )\bigr]. \]
Analogously, $Q^1 _L\ol{X}\bigl[ \varphi_1(\ol{X}), \dots,
  \varphi_{s-1}(\ol{X})\bigr]$ can be expressed as  
\[Q^{\star}_L\ol{X}\bigl[ \varphi_1(X_1/\phi_{1}(\ol{X}),\ldots, X_k/\phi_{k}(\ol{X}) ), \dots,
\varphi_{s-1}(X_1/\phi_{1}(\ol{X}),\ldots, X_k/\phi_{k}(\ol{X})
)\bigr].\eqno{\qEd}
\]\vskip6 pt

\noindent By Proposition \ref{sufflecor}, the two semantics of
second-order quantifiers coincide for all the logics (with built-in or
definable arithmetic) considered in this article.

\begin{rem}\label{mary} Let $L$ and $\CA$ be as in Definition \ref{lind2}. It is worth noting that, for $m>1$, the $m$-ary second-order quantifiers $Q^1 _L$ and $Q^{\star}_L$ can be defined  by  straightforward modifications to Definition \ref{lind2}. The $m$-ary quantifiers binds  a $k$-tuple  $\ol{X} = (X_1,\dots,X_k)$ (for some $k$) of $m$-ary second-order variables in $s-1$ many formulas.  Each $X_i$ is encoded by the bit string $s_0\cdots s_{{n^m}-1}$ with the meaning $s_j =
1$ if and only if  the $j$th tuple in the lexicographic ordering of $\{0,1,\dots,n-1\}^m$ is in $X_i$. The semantics of the $m$-ary quantifiers $Q^1_L$ and $Q^{\star}_L$ can be now defined analogously to Definition \ref{lind2}.
We use the notation 
  $Q^1_L\FO$ and  $Q^{\star}_L\FO$ for the analogues of  $\msQ_L\FO$ and $\QmodL\FO$ in which the $m$-ary quantifiers $Q^1 _L$ and $Q^{\star}_L$ are allowed for  $m\ge 1$.
\end{rem}

We end this section by showing that, in the non-monadic case, the analogue of Proposition \ref{sufflecor} holds without built-in arithmetic if $L$ has a neutral letter.  
\begin{prop}\label{sufflemary} For any $L\in\Neut$, 
  $Q^{\star}_L\FO\equiv  Q^1_L\FO$.
\end{prop}
\proof We may assume that $L$ has an  alphabet $\Sigma = (a_1, a_2, \dots,a_{s})$, where $a_s$ is a neutral letter. 

We will first show that $Q^{\star}_L\FO\le  Q^1_L\FO$. The idea of the proof is to show that a formula $\psi \in Q^{\star}_L\FO$ can be replaced by 
a formula $\psi' \in Q^{\star}_L\FO$ in which only one second-order variable with higher arity is quantified. Now, in  $\psi'$, the quantifier $Q^{\star}_L$ can be replaced by $Q^1_L$ since the difference of the two semantics only appears if more than one variable is quantified. The converse inclusion follows directly from the fact that $ Q^1_L\FO\le \LeafP(L)\equiv Q^{\star}_L\FO$ (see Theorem \ref{conn}).  

Let $\psi\in Q^{\star}_L\FO$ be of the   
 $$\psi:= Q^{\star}_L\ol{X}\bigl[ \varphi_1(\ol{X}), \varphi_2(\ol{X}), \dots,
\varphi_{s-1}(\ol{X})\bigr],$$
where $\ol{X}=(X_1,\ldots,X_k)$ is a tuple of $m$-ary second-order variables. It is straightforward to construct a formula $\psi'\in Q^{\star}_L\FO$
 $$\psi':= Q^{\star}_L R_1\bigl[\varphi'_1(R_1), \varphi'_2(R_1), \dots,
\varphi'_{s-1}(R_1)\bigr],$$
where the arity of $R_1$ is $m+\lfloor \log(k)\rfloor +1$, which is equivalent to $\psi$ over structures $\CA$ with $|\dom(\CA)|\ge2$. Let $\CA$ be a structure such that $|\dom(\CA)|\ge 2$ and $A_i\subseteq \dom(\CA)^m$.  The idea is to encode the tuple $\ol{A}=(A_1,\ldots,A_k)$ by a unique $(m+\lfloor \log(k)\rfloor +1)$-ary relation $B_{\ol{A}}$
$$ B_{\ol{A}}=\bigcup_{1\le i\le k} \{(j^i_1,\ldots,j^i_{\lfloor \log(k)\rfloor +1})\}\times A_i, $$
where $j ^i_1\cdots j^i_{\lfloor \log(k)\rfloor +1}$ is the length $\lfloor \log(k)\rfloor +1$ binary representation of $i$. This ensures that  the  ordering of the tuples $\ol{A}$ (see Definition \ref{lind2} and Remark \ref{mary}) coincides with the ordering of the corresponding codes $B_{\ol{A}}$. Therefore, it suffices to construct the formulas $\varphi'_i(R_1)$ in such a way that, for all  $A_1,\ldots,A_k \subseteq \dom(\CA)^m$  
$$\CA\models\varphi'_i(B_{\ol{A}})\iff \CA\models \varphi_i(A_1,\ldots,A_k),$$
and,  if $B\neq B_{\ol{A}}$ for all $\ol{A}$, then  $\CA\not\models \varphi'_i(B)$ implying that  the formulas $\varphi_i'(R_1)$ output the neutral letter when $R_1$ is interpreted by the relation $B$. 

In order to ensure that   
$\psi$ and $\psi'$ are equivalent also over structures $\CA$ for which $|\dom(\CA)|=1$, we may further replace the formulas 
$\varphi_i'(R_1)$ by formulas $\varphi_i^*(R_1,\ldots, R_k)$, where each  $R_i$, for $2\le i\le k$, is also $(m+\lfloor \log(k)\rfloor +1)$-ary and $\varphi^*_i(R_1,\ldots, R_k)$ has the following form
\[ (|\dom(\CA)|=1\wedge \chi_i(R_1,\ldots, R_k))\vee (|\dom(\CA)|>1 \wedge \bigwedge_{2\le i\le k} R_i=\emptyset \wedge \varphi_i'(R_1))), \]
where $\chi_i$ simulates the behavior of $\varphi_i$ on structures with cardinality $1$ (on structures with cardinality $1$ the quantifiers $Q^1_L$ and $Q^{\star}_L$ are equivalent).  Note that, for $\CA$ with $|\dom (\CA)|\ge 2$, the formulas
 $\varphi^*_i(R_1,\ldots, R_k)$ output the neutral letter if $R_i^{\CA}\neq \emptyset$ for some $2\le i\le k$. It follows that for all $\CA$
\[\CA\models \psi \Leftrightarrow \CA \models Q^1 _L \ol{R}\bigl[ \varphi^*_1(\ol{R}), \varphi^*_2(\ol{R}), \dots,
\varphi^*_{s-1}(\ol{R})\bigr]. \]

For the converse, it suffices to note that a
polynomial-time non-deterministic Turing machine with the leaf language $L$ can  easily evaluate sentences of $Q^1_L\FO$ implying that  $Q^1_L\FO\le \LeafP(L)$ (see \cite{buvo98}). Therefore, by Theorem \ref{conn}, we get that $Q^1_L\FO\le  Q^{\star}_L\FO$.
\qed

\subsection{Leaf languages}\label{leafl}
In this section we give a brief introduction to the leaf languages approach in computational complexity.

 The leaf languages approach was introduced by Bovet, Crescenzi and Silvestri in \cite{bocrsi92} and independently by Vereshchagin in \cite{ver93a}. In this approach the acceptance of a word input to a nondeterministic Turing machine depends only on the values printed at the leaves of the computation tree. 

Let $M$ be a nondeterministic Turing machine which halts on every computation path with some order on the nondeterministic choices. The order of the nondeterministic choices induces a left-to-right ordering of all the leaves in the computation tree of $M$ on input $x$. Define $\leafM(x)$ to be the concatenation of the symbols printed at the leaves of the computation tree in this order. Given now a language  $B$, the class $\LeafP(B)$ contains those languages $L$ for which there is a polynomial-time non-deterministic Turing machine $M$ such that for all inputs $x$: $x\in L$ iff  $\leafM(x)\in B$.

Let us look at some examples. Define $\Maj := \{ w\in \{0,1\}^+ |\  |w|_1>|w|_0  \} $. 
\begin{exa} The following leaf language classes are well known:
\begin{enumerate}[$\bullet$]
\item $\np=  \LeafP(0^*1(0+1)^*)$,
\item $\pp = \LeafP(\Maj)$,
\item $ \Modq = \LeafP(L_{\text{mod }q})$.
\end{enumerate}
\end{exa}
\noindent
In \cite{buvo98} complexity classes defined by  leaf languages were  logically characterized in terms of generalized second-order quantifiers. In particular, for every language $B$ that has a
 neutral letter the following was shown to hold. 
\begin{thm}[\cite{buvo98}]\label{conn}
For any $B\in\Neut$, $\LeafP(B)\equiv Q^{\star}_B\FO$.
\end{thm}
Note that, for Theorem \ref{conn} to hold, the quantifier $Q^{\star}_B$ must be
allowed to bind relation variables of arbitrary arity (see Remark \ref{mary}). 
Although the $m$-ary second-order quantifiers $Q^1 _B$ and  $Q^{\star}_B$ differ, in Theorem \ref{conn} we can equivalently use the semantics $Q^1 _B$ instead of   $Q^{\star}_B$ by Proposition \ref{sufflemary}.

Since it is known that there are regular languages $B$, e.g., the word-problem for
the group $S_5$, for which $\LeafP(B)\equiv \pspace$ \cite{helascvowa93}, we conclude that for
such $B$,
\begin{equation*}
Q^{\star}_B\FO\equiv \pspace .
\end{equation*}
\subsection{Finite leaf automata}
The automata theoretic analogue of a Turing machine with a leaf language is the so-called finite leaf automaton \cite{pevo01}.

A \emph{finite leaf automaton} is a tuple
$M=(Q,\Sigma,\delta,s,\Gamma,\beta)$ where $Q$ is a finite set of
states, $\Sigma$ is an alphabet, the input alphabet,
$\delta\colon Q\times\Sigma\rightarrow Q^+$ is the transition
function, $s\in Q$ is the  initial state, $\Gamma$ is an
alphabet, the  leaf alphabet, and $\beta\colon
Q\rightarrow\Gamma$ is a function that associates a state $q$ with its
 value $\beta(q)$.  The sequence $\delta(q,a)$, for $q\in Q$
and $a\in\Sigma$, contains all possible successor states of $M$ when
reading letter $a$ while in state $q$, and the order of letters in
that sequence defines a total order on these successor states.
This definition allows the same state to appear more than once as a
successor in $\delta(q,a)$.

Let $M$ be as above.  The computation tree $T_M(w)$ of $M$ on input
$w$ is a labeled directed rooted tree defined as follows:
\begin{enumerate}[$\bullet$]
\item The root of $T_M(w)$ is labeled $(s,w)$.
\item Let $v$ be a node in $T_M(w)$ labeled by $(q,x)$, where
$x\neq\epsilon$ (the empty word), $x=ay$ for $a\in\Sigma$,
$y\in\Sigma^*$.  Let $\delta(q,a)=q_1q_2\cdots q_k$. Then $v$ has $k$
children in $T_M(w)$, and these are labeled by
$(q_1,y),(q_2,y),\dots,(q_k,y)$ in this order.
\end{enumerate}
If we look at the tree $T_M(w)$ and attach the symbol $\beta(q)$ to a
leaf in this tree with label $(q,\varepsilon)$, then $\leafM(w)$ is
defined to be the string of symbols attached to the leaves, read from
left to right in the order induced by $\delta$.

\begin{defi}
For $A\seq\Gamma^*$, the class $\LeafFA(A)$ consists of all languages
$B\seq\Sigma^*$, for which there is a leaf automaton $M$ as just
defined, with input alphabet $\Sigma$ and leaf alphabet $\Gamma$ such
that for all $w\in\Sigma^*$, $w\in B$ iff $\leafM(w)\in A$. If $C$ is  a class of languages then
$\LeafFA(C)\equiv \cup _{A\in C}\LeafFA(A)$.
\end{defi}

In \cite{pevo01} the acceptance power of leaf automata with different kinds of leaf languages was examined.
 It was shown that,
with respect to  resource-bounded leaf language classes, there is not much difference, e.g., between automata and Turing
machines.  On the other hand, if the leaf language class is a formal language class then the differences can be huge. In particular, it was shown   that
\[\LeafFA(\REG)\equiv\REG,\]
 while it is known that
\[\LeafP(\REG)\equiv \pspace.\]
 In \cite{pevo01}
the power of  $\LeafFA(\CFL)$ was left as an open question. The only upper
and lower bounds known at that time were $\CFL\subsetneq\LeafFA(\CFL)\seq\dspace(n^2)\cap\dtime\bigl(2^{O(n)}\bigr)$.
Recently  it was  shown by Lohrey \cite{loh08} that indeed $\LeafFA(\CFL)$ does contain a $\pspace$-complete language.

In \cite{MR1908783} the class $\LeafFA(L)$ was logically characterized assuming that
the language $L$ has a neutral letter.
\begin{thm}[\cite{MR1908783}]\label{neutral}
For any $L\in\Neut$, $\LeafFA(L)\equiv \msQ_L\FO$.
\end{thm}

\begin{cor}\label{Grppspace}$\pspace$-complete languages can be defined in  $\msQGrp\FO$.
\end{cor}
\proof By the result of  \cite{loh08}, there is a language $L\in \CFL$ such that the class  $\LeafFA(L)$ contains
a   $\pspace$-complete language.  Since $L$ reduces via a length-preserving homomorphism 
to some  groupoid word-problem $A$ \cite{belemc93}, it follows that also the class  $\LeafFA(A)\equiv  \msQ_A\FO$ contains a  $\pspace$-complete language.
\qed

\subsection{Complexity theory}

We assume familiarity with the basic notions in formal languages and complexity theory, e.g., complexity classes such as $\np$, $\pp$, $\ph$, and $\pspace$. 
$\REG$ and $\CFL$ refer to the regular and context-free languages. Also, $\LOGCFL$ denotes the closure of $\CFL$ under log-space reductions.

In this article  $\AC$, $\ACC$,  $\tc$, $\NC$, and  $\SAC$ refer to the classes of languages recognized by $\DLOG$-uniform families $(C_n)_{n\in\NN}$ of  polynomial-size circuits with the following kinds of gates: 
\begin{enumerate}[$\ACC$]
\item[$\AC$:] the circuit $C_n$ may have NOT, unbounded fan-in AND and OR gates, and constant depth.
\item[$\ACC$:]  the circuit $C_n$ may have NOT, unbounded fan-in AND, OR and $\MOD$  gates, and constant depth.
\item[$\tc$:] the circuit $C_n$ may have NOT, unbounded fan-in AND, OR, and MAJORITY gates, and constant depth.
\item[$\NC$:]  the circuit $C_n$ may have NOT,  bounded fan-in AND and OR gates, and $O(\log(n))$ depth.
\item[$\SAC$:]  the circuit $C_n$ may have input level NOT gates,  bounded fan-in AND and unbounded fan-in OR gates, and $O(\log(n))$ depth.
\end{enumerate}
 The requirement of $\DLOG$-uniformity means that $(C_n)_{n\in\NN}$, as a family of  directed acyclic graphs, can be recognized by a deterministic Turing machine, with random access to its input,  in time $O(\log(n))$ (see \cite{vol99} for details). Note that, e.g., the classes ${\rm AC}^i$ and ${\rm TC}^i$ are defined analogously as above but allowing $O(\log^i(n))$ circuit-depth. 

In this article we also discuss certain complexity classes defined in terms of alternating Turing machines.  We denote by $\ATIME(t(n))$, the class of languages which can be recognized in time $t(n)$  by some  alternating Turing machine.  

Let $M$ be an alternating Turing machine accepting $x$ and denote by $T$ the computation tree  produced by $M$ with input $x$. An \emph{accepting computation subtree} $S$ of $M$ on input $x$ is a subtree of  $T$ witnessing that $M$ accepts $x$. The idea is that all the nodes in $S$ must be accepting configurations, and, furthermore, $S$ must contain the initial configuration, i.e., the  root of $T$, all successors of universal configurations, and exactly one successor of each existential configuration.  

We say that an alternating machine $M$ is \emph{tree-size bounded} by $t\colon \mathbb{N}\rightarrow \mathbb{N}$ if for every $x$ accepted by $M$ there is an accepting computation subtree of $M$ on input $x$ which has at most $t(|x|)$ nodes. 
Let now 
 \[\ASPACETS(s(n),t(n))\]
 denote the class of languages which can be recognized by an  alternating Turing machine $M$ which is  space bounded by $s$ and tree-size bounded by $t$. 

The following (non-trivial)  inclusions and  equalities are known to hold among the classes defined above: 
\begin{eqnarray*}
  \tc \subseteq \NC=\ATIME(\log(n)) \subseteq \SAC &=& \LOGCFL\\ 
&=& \ASPACETS(\log(n),n^{O(1)}).   
\end{eqnarray*}
The last two equalities where shown by Venkateswaran  \cite{ven91} and  Ruzzo \cite{ruz80}, respectively.

\section{Groupoidal quantifiers}

\noindent In this section we consider second-order monadic groupoidal
quantifiers. We show that the extension of $\SOM$ in terms of all
second-order monadic groupoidal quantifiers collapses in expressive
power to its fragment $\Qmod\FO$ (respectively to
$\msQGrp\FO(+,\times)$).

The following result on first-order groupoidal quantifiers will be central for our reasoning.
Below, $\qfree$ refers to the set of quantifier-free formulas (of suitable signature) in which the predicates $+$ and $\times$ do not appear.

\begin{thm}[\cite{lamcscvo01}]\label{grup} $\QGrp \qfree \equiv  \FO(\QGrp)\equiv   \FO(\QGrp, +,\times)\equiv\LOGCFL.$ 
\end{thm}
We shall use the following  version  of Theorem \ref{grup}.

\begin{lem}\label{new}Let $\tau=\{c_1,\ldots,c_s\}$, where $c_1,\ldots,c_s$ are  constant symbols. Then on $\tau$-structures
\begin{equation*}
\QGrp \qfree \equiv  \FO(\QGrp)\equiv \FO(\QGrp, +,\times)  .
\end{equation*}
\end{lem}
\proof The idea of the proof is to translate $\varphi \in \FO(\QGrp, +,\times)[\tau]$ into $\varphi^*\in  \FO(\QGrp, +,\times)$ of a suitable string signature using a simple encoding of $\tau$-structures into strings. By Theorem \ref{grup}, we may then replace $\varphi^*$ by an equivalent formula $\theta\in \QGrp \qfree$. Finally, we show that $\theta$ can be translated back to a formula
$\theta^*\in \QGrp \qfree[\tau]$ in such  a way that  $\theta^*$ and $\varphi$ are equivalent.
 
Suppose that $K$ is a class of $\tau$-structures definable by $\varphi\in\FO(\QGrp, +,\times)$.
We shall encode $K$ as a class of strings over signature
$\langle P_{S_1},\ldots, P_{S_{2^s}}\rangle$, where $S_1,\ldots,S_{2^s}$ is some fixed enumeration of
the subsets of $\{ c_1,\ldots,c_s\}$. We associate every $\tau$-structure $\CA$ with a unique string
$w_{\CA}$ over the same universe in the following way. For $S\subseteq \{ c_1,\ldots,c_s\}$,  define
$$P^{w_{\CA}}_S=\{b\ |\  c_i^{\CA}=b \Leftrightarrow c_i\in S\}.$$
Note that the predicate $P_{\emptyset}$ is interpreted by the set $\{0,\ldots,n-1\}\setminus \{c_1^{\CA},\ldots,c_s^{\CA}\}$ where $\{0,\ldots,n-1\}$ is the universe of $\CA$.

 Let $\varphi^*$ be acquired from $\varphi$ by replacing atomic subformulas
of the form $c_i=t$ by $\vee_{c_i\in S}P_S(t)$ and $c_i=c_j$ by the formula $\exists y (\vee_{c_i,c_j\in S}P_S(y))$.
It is now obvious how to translate atomic formulas using the predicates $+,\times$, and $<$, e.g., the formula $c_i<t$ is replaced by $\exists y(\vee_{c_i\in S}P_S(y))\wedge y<t)$. It is easy to show using induction on the construction of $\varphi\in\FO(\QGrp, +,\times)$ that for all $\tau$-structures
 $\CA$, 
\[ \CA\models \varphi \Leftrightarrow  w_{\CA}\models \varphi^*.\]
By Theorem \ref{grup} there is a sentence $\theta\in \QGrp \qfree$ which is equivalent to $\varphi^*$
over strings. Let $\theta^*$ be acquired from $\theta$ by replacing subformulas  $P_{S}(t)$  by
\[(\bigwedge_{c_i\in S} c_i=t)\wedge (\bigwedge_{c_j\in \{c_1,\ldots,c_s\}\setminus S} c_j\neq t).\]
 Again by induction on $\theta\in \QGrp \qfree$  we get that for all $\tau$-structures
 $\CA$,
\[\CA\models \theta^* \Leftrightarrow  w_{\CA}\models \theta.\]
 It follows that $\theta^*\in \QGrp \qfree$ defines $K$.
\qed

We are now ready for the main result of this section. Note that the built-in predicates  $+$ and $\times$ are definable 
already in terms of (first-order) majority quantifiers (see  \cite{baimst90} and \cite{lamcscvo01}) and hence definable in the logics  in which groupoidal quantifies are allowed to be nested.

\begin{thm}\label{11}  $\SOM(\Qmod)\equiv \Qmod\FO$.
\end{thm}
\proof 
Fix a signature $\tau =\langle P_{a_1},\dots,P_{a_s}\rangle$.
Suppose that  $\varphi\in \SOM(\Qmod)$ is a sentence. We will show how to construct a sentence  of the logic 
$ \Qmod\FO$  equivalent to $\varphi$. The idea of the proof is to represent 
 $\varphi \in \SOM(\Qmod)$, and the  language of signature $\tau$ defined by $\varphi$,  in terms of  $\varphi^*\in \FO(\QGrp,+,\times)$, and the class of $\sigma$-structures defined by $\varphi^*$,  where  $\sigma=\{ c_1,\ldots,c_s\}$ and $c_1,\ldots, c_s$ are constant symbols. More precisely, by representing  $\tau$-structures of cardinality $n$ by $\sigma$-structures of cardinality $2^n$, we can replace second-order variables over the domain $\{0,\ldots,n-1\}$ by first-order variables ranging over $\{0,\ldots,2^n -1\}$ using the $\bit$-predicate. Then we apply Lemma \ref{new} to get a formula $\theta \in \QGrp \qfree $    equivalent to $\varphi^*$. Finally, we show that $\theta$ can be translated back to a formula  $\theta '\in \Qmod\FO[\tau]$ in such a way that the original formula $\varphi$ and $\theta'$ are equivalent.  

Denote by $\sigma=\{c_1,\ldots,c_s\}$ the signature where each $c_i$ is a constant symbol. For a $\tau$-structure
$\CA=\langle \{0,\ldots,n-1\},<,P_{a_1} ^{\CA},\ldots, P_{a_s} ^{\CA}\rangle$, let $\CA^*$ be the following $\sigma$-structure
\[\CA^* = \langle \{0,\ldots,2^n-1\},<,+,\times,c_1^{\CA^*},\ldots, c_s^{\CA^*}\rangle,  \]
where $c_i^{\CA^*}$ is the unique integer ($<2^n$) whose binary representation is given by
$s_0\cdots s_{n-1}$ where  $s_j =1 \iff j\in P_{a_i} ^{\CA}$.

We shall first show that there is a sentence $\varphi^*\in \FO(\QGrp,+,\times)[\sigma]$ such that
for all $\tau$-structures $\CA$,
\begin{equation}\label{translation}
\CA \models \varphi \Leftrightarrow  \CA^*\models \varphi^*  . 
\end{equation}
The translation $\varphi \rightsquigarrow\varphi^*$  is defined inductively as follows.  For $\varphi$ of the form $x_i=x_j$ or $x_i<x_j$, $\varphi^* := \varphi$, and in the remaining  cases (we may exclude the definable  constants $\mi$,  $\ma$, and the  second-order existential quantifier  from $\SOM(\Qmod)$ since $Q^{\star}_{L_{\exists}}$ is available)  the translation is defined in the following way:
\begin{eqnarray*}
P_{a_i}(x_i) &\rightsquigarrow& \bit(c_i,n-(x_i+1))\\
Y_i(x_j) &\rightsquigarrow& \bit (y_i,n-(x_j+1))\\
\psi\wedge \phi &\rightsquigarrow& \psi^*\wedge \phi^* \\
\neg \psi &\rightsquigarrow& \neg \psi^*\\
\exists x_i \psi &\rightsquigarrow& \exists x_i(x_i<n \wedge \psi^*(x_i))\\
Q^{\star}_L Y_1,\ldots,Y_k[\psi_1,\ldots,\psi_{s-1}] &\rightsquigarrow& Q_{L}y_1,\ldots,y_k[\psi_1 ^*,\ldots,\psi^* _{s-1}]
\end{eqnarray*}
It is straightforward to show using induction on the construction of $\varphi\in  \SOM(\Qmod)[\tau]$, that for all  $\tau$-structures $\CA$ and assignments $s$,
\[\CA \models_s \varphi \Leftrightarrow  \CA^*\models_{s^*} \varphi^* , \] 
where the  assignment $s^*$ over $\CA^*$ is defined such that  $s^*(x_i)=s(x_i)$, for all first-order variables $x_i$, and, for a second-order $Y_i$: if $s(Y_i)=A\subseteq \{0,\ldots,n-1\}$, then $s^*(y_i)$ is the unique $a<2^n$ whose length $n$ binary representation is given by
$s_0\cdots s_{n-1}$ where  $s_j =1 \iff j\in A$.  

Above, we use the predicate  $\bit$ which is definable in $\FO(+,\times)$ (see, e.g., \cite{imm99}). Note also that, using   the predicate $\bit$,  the integer $n$ can be  easily defined over the structure $\CA^*$.
 
By Lemma \ref{new}, there is a sentence
\[ \theta = Q_L x_1,\ldots,x_l [\chi_1,\ldots, \chi_{w}],\]
where each $\chi_i$ is quantifier-free and does not contain the predicates $+$ and $\times$, equivalent to $\varphi^*$.
The idea is now to translate $\theta$ into  $\theta'\in \Qmod\FO$
by  changing first-order variables to second-order variables. Denote by $X=Y$ the formula
$\forall z(X(z)\leftrightarrow Y(z))$,
and by $X<Y$ the first-order formula defining the ordering of subsets when treated  as length $n$ binary strings. The translation  $\theta \rightsquigarrow\theta'$
is now defined by
\begin{eqnarray*}
t=\hat{t} &\rightsquigarrow& X_{t}=X_{\hat{t}}\\
t<\hat{t} &\rightsquigarrow&  X_{t}<X_{\hat{t}}\\
\psi\wedge \phi &\rightsquigarrow& \psi'\wedge \phi' \\
\neg \psi &\rightsquigarrow& \neg \psi'\\
Q_L x_1,\ldots,x_v[\psi_1,\ldots,\psi_{v}] &\rightsquigarrow& Q^{\star}_{L}X_1,\ldots,X_v[\psi_1' ,\ldots,\psi'_v]
\end{eqnarray*}
Above, $t$  is  either min, max, $c_l$, for $1\le l\le s$, or a variable $x$, and, respectively, $X_{t_i}$ is  either $\bot$, $\top$, $P_{a_l}$, or $X$. A straightforward  induction implies, in particular, that for all sentences $\psi \in\QGrp \qfree[\sigma]$, and $\tau$-structures $\CA$
\[\CA \models \psi' \Leftrightarrow  \CA^*\models\psi, \]
where $ \CA^*$  is defined as above. It is now immediate that $\theta'$ and the original  sentence $\varphi \in \SOM(\Qmod)[\tau]$ are equivalent.
\qed

Note that, by Proposition \ref{sufflecor},  we do not need to consider the semantics $Q^1_L$ separately. By combining 
Theorem \ref{11} and Proposition \ref{sufflecor}, we get
\begin{cor}\label{cor}
\begin{align*}\SOM(\msQGrp) &\equiv \msQGrp\FO(+,\times)\\
 &\equiv  \Qmod\FO \equiv \SOM(\Qmod).
\end{align*}
\end{cor}\vskip3 pt

\noindent We close this section by showing that a much stronger analogue of Corollary \ref{cor} holds. Recall that the so-called Greibach's hardest context-free language $H$ is a  nondeterministic
version of the Dyck language $D_2$, the language of all syntactically correct sequences consisting of letters for two types of parentheses. It is known that every $L\in \CFL$ reduces to $H$ under some homomorphism \cite{gre73}. It was shown in \cite{lamcscvo01} that the statement of Theorem \ref{grup} remains valid even if the
logic $\QGrp \qfree$ is replaced by the logic $Q_{{\rm pad}(H)}\qfree$, where ${\rm pad}(H)$ is $H$ extended by a neutral letter. This result directly implies the following strengthening of Corollary \ref{cor}.

\begin{thm}\label{Greibach}
\begin{align*}\SOM(\msQGrp) &\equiv  \msQH\FO(+,\times)\\
 &\equiv  \QmodH \FO \equiv \SOM(\Qmod).
\end{align*}
\end{thm}
\proof The proof is analogous to the proof of Theorem \ref{11}. It suffices to prove the last equality in the statement of the theorem. Suppose that  $\varphi\in \SOM(\Qmod)$ is a sentence. By an analogous argument as in the proof of Theorem \ref{11}, we first translate
$\varphi$ into a sentence $\varphi^* \in\FO(\QGrp,+,\times)$. Then we replace $\varphi^*$ by an equivalent sentence $\theta\in Q_{{\rm pad}(H)}\qfree$. Now, again by an analogous argument as in the proof of Theorem \ref{11}, $\theta$ can be translated back to the logic  $\QmodH \FO$.\qed

\section{Monoidal quantifiers}

\noindent In this section we consider second-order monadic monoidal
quantifiers.

As already mentioned, the following result completely characterizes the picture in the case of the semantics $Q^1 _L$ without built-in arithmetic.
\begin{thm}[\cite{MR1908783}]\label{monoids}
$\msQMon\FO \equiv \SOM(\msQMon) \equiv \REG\equiv \exists\SOM.$
\end{thm}
Interestingly, the expressive power of  monoidal quantifiers collapses to regular languages when  built-in arithmetic is not present. Under reasonable complexity theoretic assumptions,  the corresponding equalities between the logics in Theorem \ref{monoids} do not hold with built-in arithmetic. Furthermore, it is an open question if the analogue of Theorem \ref{monoids} holds with respect  to the semantics $Q^{\star}_L$. Again, by Proposition \ref{sufflecor}, we however know that the  semantics coincide assuming built-in arithmetic. 

\begin{thm}\label{MonoidswA} The following equivalences hold
\begin{enumerate}[\em(1)]
\item\label{MonoidswA1} $\msQMon\FO(+,\times) \equiv \QmodM\FO(+,\times)$,
\item\label{MonoidswA2} $\SOM(\msQMon,+,\times) \equiv \SOM(\QmodM, +,\times)$.
\end{enumerate}
\end{thm}
\noindent
Note that also in equivalence \ref{MonoidswA2} of  Theorem \ref{MonoidswA}  the arithmetic predicates (in fact $+$ would also suffice) need to be assumed by Theorem \ref{monoids}, i.e., $+$ and $\times$  are not definable  in $\SOM(\msQMon)$.  It is an open question whether the equivalences of Theorem \ref{MonoidswA} hold without built-in arithmetic.

\section{Complexity results}
\noindent In this section we study the data complexity of the logics
discussed in the previous sections.

We begin with a simple logical padding argument which allows us to utilize Theorem \ref{conn} in the context of second-order monadic quantifiers. Recall  that, in the statement of  Theorem \ref{conn}, the  quantifier $Q^{\star}_B$ is allowed to bind relation variables of arbitrary arity (see Remark \ref{mary}).  Below, we do not distinguish notationally between a string $w$ of alphabet $(a_1,\dots,a_s)$,  and the string structure of signature $\langle P_{a_1},\dots,P_{a_s}\rangle$ corresponding to $w$.   
\begin{prop}\label{padd}Let $L$ be a language and suppose that a language $A$ of alphabet  $\Sigma$ is definable by a sentence $\varphi\in Q^{\star}_L\FO$.
Let $k$ be the arity of the relations quantified in $\varphi$ and $\sharp\notin \Sigma$. Then the language
\[A^* =\{ w^{\smallfrown}\sharp^{|w|^k-|w|}\ |\ w\in A\}\]
is definable in $\FO(\QmodL,+,\times)$.
\end{prop}
\proof 
Let $\varphi$ be of the form
\[Q^{\star}_L R_1,\ldots,R_t [\psi_1,\ldots,\psi_s], \]
where each of the relations $R_i$ has arity $k$.  Define a translation  $\varphi \rightsquigarrow\varphi^*$   as follows. For $\varphi$ of the form $x_i=x_j$,  $x_i<x_j$, or $P_{a_i}(x_j)$, $\varphi^* := \varphi$, and in the remaining  cases (we exclude the definable constants $\mi$ and $\max$) the translation is defined in the following way:
\begin{eqnarray*}
R_i(x_1,\ldots,x_k) &\rightsquigarrow& \exists z(X_{R_i}(z)\wedge z=|w|^{k-1}x_1+\cdots+|w|x_{k-1} + x_k)  \\
\psi\wedge \phi &\rightsquigarrow& \psi^*\wedge \phi^* \\
\neg \psi &\rightsquigarrow& \neg \psi^*\\
\exists x \psi &\rightsquigarrow& \exists x(x< |w| \wedge \psi^*(x))\\
Q^{\star}_L R_1,\ldots,R_t[\psi_1,\ldots,\psi_{s-1}] &\rightsquigarrow& Q^{\star}_LX_{R_1},\ldots,X_{R_t}[\psi_1 ^*,\ldots,\psi^* _{s-1}]
\end{eqnarray*}
It is straightforward to show using induction on  $\psi\in Q^{\star}_L\FO$ that for all $w$ and assignments $s$
\[ w\models_s \psi \Leftrightarrow  w^{\smallfrown}\sharp^{|w|^k-|w|}\models_{s^*} \psi^*,   \]
where $s^*$ agrees with $s$ with respect to first-order variables, and 
\[s^*(X_{R_i})=\{ a\ |\ a=|w|^{k-1}b_1+\cdots+|w|b_{k-1} + b_k  \textrm{ for some }(b_1,\ldots,b_k)\in s(R_i) \}. \]
 Note also that, by our conventions, the universe of  $w^{\smallfrown}\sharp^{|w|^k-|w|}$ is  $\{0,\ldots, n^k-1\}$ hence there is 1-1 correspondence between the subsets of 
$\{0,\ldots, n^k-1\}$ and the $k$-ary relations over the universe, $\{0,\ldots, n-1\}$, of $w$.

Finally, the language $A^*$ is defined by $\varphi^*\wedge\chi$, where $\chi\in \FO(+,\times)$ and 
\[w\models \chi \Leftrightarrow \exists \tilde{w}(w=\tilde{w}^{\smallfrown}\sharp^{|\tilde{w}|^k-|\tilde{w}|}).\eqno{\qEd}  \]

\noindent Proposition \ref{padd} shows that logics
$\FO(\QmodL,+,\times)$ can be quite powerful. In fact, it is apparent
from the proof that if, e.g., $\varphi \in Q^{\star}_L\FO$ in the
proof of Proposition \ref{padd} defines a $\pspace$-complete language,
then the language defined by $\varphi^*\in \QmodL\FO(+,\times)$ is
also $\pspace$-complete.

\begin{cor}\label{monpspace} In the logic $\msQMon\FO(+,\times)$,  $\pspace$-complete languages can be defined.  
\end{cor}
\proof  This follows, e.g., by the fact that 
\begin{equation*}
Q^{\star}_B\FO\equiv \pspace,
\end{equation*}
where $B$ is the word-problem for the group $S_5$ (see Section \ref{leafl}), and by Proposition \ref{sufflecor}.
\qed\vskip3 pt

\noindent Recall that in the case of groupoidal quantifiers, already
in $\msQGrp\FO$ $\pspace$-complete languages can be defined by
Corollary \ref{Grppspace}.

Next we show that the logics $ \SOM(\QmodM,+,\times)$ and $\SOM(\Qmod)$ capture the exponential versions of the language classes captured by the logics  $\FO(\QMon,+,\times)$ and $\FO(\QGrp,+,\times)$. As already noted in Theorem \ref{grup},  the logic $\FO(\QGrp,+,\times)$ corresponds to  $\LOGCFL$ \cite{lamcscvo01}. On the other hand, in  \cite{baimst90} it was show that 
\begin{equation}\label{ATIME(logn)}
 \FO(\QMon,+,\times) \equiv \NC = \ATIME(\log(n)).
\end{equation}
\begin{defi} For $n\in \NN$, denote by $\bin(n)$ the binary representation of $n$ without leading zeros. Let  $L\subseteq \{0,1\}^+$ and $1L=\{1w\ |\ w\in L \}$. Define now $\tally(L)$ as
 \[\tally(L)=\{ 1^n\ |\ \bin(n)\in 1L \}. \] 
\end{defi}\vskip6 pt

\noindent Let us now define the classes of languages
$2^{\ATIME(\log(n))}$ and $2^{\LOGCFL}$ by
\begin{eqnarray*}
2^{\ATIME(\log(n))} &=& \{L\subseteq \{0,1\}^+|\ \tally(L)\in \ATIME(\log(n)) \},\\
2^{\LOGCFL} &=& \{L\subseteq \{0,1\}^+|\ \tally(L)\in \LOGCFL\}.
\end{eqnarray*}
The following is easily seen to hold:
\begin{prop}\label{easy}The following equalities hold
\begin{enumerate}[\em(1)]
\item $2^{\ATIME(\log(n))}=\ATIME(n)$,
\item $2^{\LOGCFL}= \ASPACETS(n,2^{O(n)})$.
\end{enumerate}
\end{prop}
\proof
The first equality is obvious and the second follows from Ruzzo's characterization  of $\LOGCFL$: 
\[\LOGCFL= \ASPACETS(\log(n),n^{O(1)})\]
  (see  \cite{ruz80} and \cite{vol99}).
\qed

\begin{rem} By the above, we immediately get that  $$\NSPACE(n)\subseteq 2^{\LOGCFL}.$$   It is also straightforward to show that  $2^{\LOGCFL}$ includes the languages that can be recognized in  linear time on a Threshold Turing machine (introduced in  \cite{pasc88}).
\end{rem}

The main result of  this section can be now stated as follows:
\begin{thm}\label{compres}The following equivalences hold
\begin{enumerate}[\em(1)]
\item\label{main1} $\SOM(\QmodM,+,\times)\equiv \ATIME(n)$, 
\item\label{main2} $\SOM(\Qmod) \equiv 2^{\LOGCFL}.$
\end{enumerate}
\end{thm}
\proof We will prove equivalence \ref{main2}. Equivalence \ref{main1} is proved  analogously using the fact that 
 $\ATIME(\log(n))\equiv \FO(\QMon,+,\times)$ (see  \eqref{ATIME(logn)}).

 We will show that,  for all $L\subseteq \{0,1\}^+$,  $L$ is definable in $\SOM(\Qmod)$  iff $\tally(L)\in \LOGCFL$. Since $ \LOGCFL\equiv \FO(\QGrp,+,\times)$, it suffices to show that for all $L\subseteq \{0,1\}^+$,  $L$ is definable in $\SOM(\Qmod)$  iff $\tally(L)$ is definable in  $\FO(\QGrp,+,\times)$.

We will first
show that if $L$ is definable in  $\SOM(\Qmod)$, then $\tally(L)$ can be defined in $\FO(\QGrp,+,\times)$.
The idea is now to translate formulas between  string structures 
\begin{equation}\label{w}
w=\langle\{0,\ldots, m-1\},<,P_1,P_0\rangle,
\end{equation}
  and
\begin{equation}\label{n}
1^n=\langle\{0,\ldots, n-1\},P_1, <,+,\times\rangle,
\end{equation}
  where  $w\in \{1,0\}^+$, $\bin(n)=1w$, and $P_1=\{0,\ldots, n-1\}$. Some technical difficulties arise here, which were not encountered  in the proof of  Theorem \ref{11}, due to the fact that the sizes of the universes of  $w$ and $1^n$ are not necessarily exactly of the form $l$ and $2^l$ for some $l$. 

We define a translation  $\varphi \rightsquigarrow\varphi^*$  of  $\varphi \in\SOM(\Qmod)$  into  $\varphi^* \in \FO(\QGrp,+,\times)$ below. 
Analogously to the  proof of  Theorem \ref{11},  it can be shown using  induction  on $\varphi\in  \SOM(\Qmod)$,  that for all  $w$ and assignments $s$,
\[w \models_s \varphi \Leftrightarrow  1^n\models_{s^*} \varphi^* . \] 
The assignment $s^*$ is defined  so that it agrees with $s$ on first-order variables $x_i$, and, for a variable $y_i$, corresponding to a second-order variable  $Y_i$, $s^*(y_i)=a<2^{m}$, where $a=\Sigma_{i=0}^{m-1}s_i2^{m-1-i}$ and $s_i=1$ iff $i\in s(Y_i)$. 

The translation $\varphi \rightsquigarrow\varphi^*$  is defined inductively as follows.
For $\varphi$ of the form $x_i=x_j$ or $x_i<x_j$, $\varphi^* := \varphi$, and in the remaining  cases (again, we exclude the definable constants $\mi$, $\max$, and the  second-order existential quantifier from $\SOM(\Qmod)$ since $Q^{\star}_{L_{\exists}}$ is available) the translation  is defined in the following way  (recall that by  \eqref{w} and \eqref{n} we have  $\lfloor \log(n)\rfloor=m=|w|$):
\begin{eqnarray*}
P_1(x_i) &\rightsquigarrow& \bit(n,\lfloor \log(n)\rfloor -(x_i+1))\\
P_0(x_i) &\rightsquigarrow& \neg \bit(n,\lfloor \log(n)\rfloor -( x_i+1))\\
Y_i(x_j) &\rightsquigarrow& \bit (y_i,\lfloor \log(n)\rfloor -( x_j+1))\\
\psi\wedge \phi &\rightsquigarrow& \psi^*\wedge \phi^* \\
\neg \psi &\rightsquigarrow& \neg \psi^*\\
\exists x_i \psi &\rightsquigarrow& \exists x_i(x_i<\lfloor \log(n)\rfloor \wedge \psi^*(x_i))\\
Q^{\star}_L Y_1,\ldots,Y_k[\psi_1,\ldots,\psi_{s-1}] &\rightsquigarrow& Q_{L}y_1,\ldots,y_k[\chi\wedge\psi_1 ^*,\ldots,\chi\wedge \psi^* _{s-1}]
\end{eqnarray*}
Note that, e.g.,  the formula $\bit(n,\lfloor \log(n)\rfloor -( x_i+1))$ above can be easily constructed even though  the integer $n$ 
is not in the universe of the structure $1^n$.   Without loss of generality,    we may  assume that the letter $a_s$ in the alphabet  $ (a_1, a_2, \dots,a_{s})$ of $L$ is a neutral letter. Now 
the formula $\chi$, used to translate $Q^{\star}_L$, ensures that the interpretation $b_1,\ldots,b_k\in \{0,\ldots, n-1\}$ of the tuple $y_1,\ldots, y_k$ does correspond to some tuple of unary relations $B_1,\ldots,B_k\subseteq \{0,\ldots,m-1\}$. The problem is that there can be more tuples $\overline{b}$ than tuples $\overline{B}$. The formula $\chi$ is defined as
\[\bigwedge _{1\le i\le k}y_i< 2^{\lfloor \log(n)\rfloor}-1.\]
Now if $\chi$ is not satisfied by $\overline{b}$, then none of the formulas $\chi\wedge\psi_i^*$
will be satisfied and hence these formulas produce the neutral letter $a_s$ when $\overline{y}$ is interpreted as $\overline{b}$.

We conclude that, by the above,  if $L$ is defined by a sentence $\varphi \in \SOM(\Qmod)$, then the  sentence 
\[(\varphi^*\wedge \forall xP_1(x))\in \FO(\QGrp,+,\times), \]
defines $\tally(L)$.

We will next define a formula translation  $\varphi\rightsquigarrow \varphi'$  mapping  $\varphi\in \FO(\QGrp,+,\times)$ into $ \varphi'\in \SOM(\Qmod)$. Again,
an analogous induction on the construction of $\varphi$ shows, in particular, that  for all sentences $\varphi$ and all $n\in \NN$, 
\[\bin(n) \models \varphi' \Leftrightarrow  1^n \models \varphi. \]
The translation $\varphi\rightsquigarrow \varphi'$  is defined by replacing first-order variables by unary second-order variables:
\begin{eqnarray*}
x=y &\rightsquigarrow& X=Y\\
x<y &\rightsquigarrow& X<Y\\
x+y=z&\rightsquigarrow& X+Y=Z\\
x\times y=z&\rightsquigarrow& X\times Y=Z\\
P_1(x)&\rightsquigarrow& \top \\
\psi\wedge \phi &\rightsquigarrow& \psi'\wedge \phi' \\
\neg \psi &\rightsquigarrow& \neg \psi'\\
\exists x \psi  &\rightsquigarrow& \exists X(\delta\wedge \psi')\\
Q_L x_1,\ldots,x_v[\psi_1,\ldots,\psi_{v}] &\rightsquigarrow& Q^{\star}_{L}X_1,\ldots,X_v[\delta \wedge\psi_1' ,\ldots,\delta\wedge \psi'_v]
\end{eqnarray*}
Above,  $X=Y$  denotes the formula $\forall z(X(z)\leftrightarrow Y(z))$. Also  $X<Y$,  $X+Y=Z$, and $ X\times Y=Z$   are  formulas defining the ordering, addition, and multiplication of unary relations,  when treated  as binary strings. Finally,
the formula $\delta$ is simply
\[\delta= X< P_1,\]
and it has an analogous role here as $\chi$ had above. Again, we may assume
that $a_s$ is a neutral letter of $L$ when translating the quantifier $Q_L$.

By the above, it holds that  for all $L\subseteq \{0,1\}^+$: if $\tally(L)\in \LOGCFL$, then $1L$ is definable in $\SOM(\Qmod)$. In order to complete the proof, it suffices to show that   $L$ is  definable in $\SOM(\Qmod)$
iff   $1L$ is  definable in $\SOM(\Qmod)$.  Note that on the computational side, $L$ and $1L$ are easily definable from each other. On the logical side, it follows from the fact that  $\SOM(\Qmod)$ is  closed under logical reductions  for which the target structure $w^*$  has size linear in $w$. The idea is that if  $|w^*|=k|w|$ then  subsets of $w^*$ can be encoded by  $k$ subsets over $ w$ and, hence, such a formula translation can be defined in terms of second-order monadic quantifiers. 
 This implies\footnote{More generally, it also implies that $\SOM(\Qmod)$ captures $2^{\LOGCFL}$ over all string signatures, since a string $w$ of any signature can be encoded in binary with length $O(|w|)$.}, in particular, that for any sentence $\varphi\in \SOM(\Qmod)$ we can construct a sentence $\varphi^*\in \SOM(\Qmod)$ which holds over $w$ iff $1w\models \varphi$. We do not give the proof here but refer to the proof of Corollary 8.6 in \cite{Kon3} in which an analogous result is proved for the extension of $\FO$ in terms of the second-order monadic majority quantifier.
\qed

Finally, we turn to the case of symmetric (commutative) languages. It is obvious that, for a symmetric language $L$, the quantifiers  $Q^1_L$ and $Q^*_L$ are equivalent. Let $\CFL^{s}$ and $\REG^{s}$ denote the classes of symmetric context-free and regular languages, respectively. 

Denote by $\lmodPH$ the linear analogue of the class $\modPH$. Recall that  $\modPH$  is the oracle hierarchy, analogous to the polynomial hierarchy $\ph$, in which the building block of the hierarchy is  $\np\cup (\Modq)_{q>1}$ and the  $(k+1)th$ level is defined by allowing access to oracles from the Boolean closure of the $k$th level. Similarly, we denote by $\lCH$ the linear analogue of the counting hierarchy $\CH$, which is the oracle hierarchy with $\pp$ as the building block. 

The expressive power of second-order monadic quantifiers defined by symmetric regular and context-free languages can be  characterized as follows:  
\begin{thm}\label{symm}The following equivalences hold
\begin{enumerate}[\em(1)]
 \item\label{symm1} $\SOM(\QmodMS,+,\times) \equiv \lmodPH,$
\item\label{symm2} $\SOM(\QmodS) \equiv \lCH.$
\end{enumerate}
\end{thm}
\proof Let us first show equivalence \ref{symm2}. Note that by Parikh's theorem on context-free languages, every symmetric context-free language is already in $\tc$ and, by \cite{baimst90}, $\tc \equiv \FO(Q_{\Maj}, +, \times)$. Therefore, we get that 
\[ \SOM(\QmodS) \equiv \SOM(\QmodMaj)  \]
by an analogous argument as in Theorem \ref{11}. In \cite{Kon3} it was shown that 
\[ \SOM(\QmodMaj)\equiv \lCH,\]
hence the claim follows. 

For equivalence \ref{symm1}, note that $\FO(Q_{\REG^s}, +, \times)\equiv \ACC$ (see \cite{baimst90}) and that, analogously to  $\AC$  and $\ph$,  $\ACC$ is the logarithmic analogue of $\modPH$ (see \cite{algo94,all99}).  Hence, by standard padding  we get that $2^{\ACC}= \lmodPH$ and, mimicking the proof of Theorem \ref{compres}, it follows that  
\[\SOM(\QmodMS,+,\times)\equiv \lmodPH.\eqno{\qEd}\]\vskip6 pt

\noindent We conclude this section by the following table summarising the results on the  data-complexity of the logics studied in this paper. 
\begin{table}[h]
\caption{Summary of the results}
\centering
\begin{tabular}{|c| c| c|c| }
\hline \hline
Logic \& built-ins & $\{\le\}$ & $\{+,\times\}$ & Result  \\
\hline \hline
 $\msQMon\FO$ & $\REG$ & $\pspace$-comp. & Thm \ref{monoids} \cite{MR1908783}, Cor \ref{monpspace} \\

 $\QmodM \FO$ & $\ge \REG$ & $\pspace$-comp. &\cite{bue62, tra61}, Cor \ref{monpspace}  \\
\hline
$\SOM(\msQMon)$  & $\REG$  & $\ATIME(n)$ & Thm \ref{monoids} \cite{MR1908783},  Thm \ref{compres}   \\

$\SOM(\QmodM)$   & $\ge \REG$  & $\ATIME(n)$  & \cite{bue62, tra61},  Thm \ref{compres}  \\
\hline
 $\msQGrp\FO$ & $\pspace$-comp. &   $2^{\LOGCFL}$ &  Cor \ref{Grppspace} \cite{loh08},   Thm \ref{compres} \\

 $\Qmod \FO$ & $2^{\LOGCFL}$ &$2^{\LOGCFL}$ &  Thm \ref{compres}    \\
\hline
$\SOM(\msQGrp)$ & $2^{\LOGCFL}$ &$2^{\LOGCFL}$ &   Thm \ref{compres}   \\

$\SOM(\Qmod)$ & $2^{\LOGCFL}$ &$2^{\LOGCFL}$ & Thm \ref{compres}    \\
\hline
$ \SOM(\QmodMS)$ & $\REG$ & $\lmodPH$ &  Thm \ref{monoids} \cite{MR1908783}, Thm \ref{symm}   \\

$\SOM(\QmodS)$ & $ \lCH$ &$ \lCH$ &Thm \ref{symm}    \\
\hline 
\end{tabular}
\label{table:summary}
\end{table}%\vspace{-6 pt}

\section{Conclusion}
\noindent We conclude with two questions for further study.
The main open question regarding  groupoidal quantifiers is  to determine whether the two variants of
semantics for second-order groupoidal quantifiers coincide in the most restricted case studied in this paper, i.e., is it the case that
\[\msQGrp\FO \equiv \Qmod\FO?\]
A positive answer would imply that
\[ \LeafFA(\CFL)= \ASPACETS(n,2^{O(n)}).\]
This would strengthen the recent $\pspace$-hardness result \cite{loh08} considerably (showing that $\LeafFA(\CFL)$ contains $\pspace$-complete problems, and answering the open question from \cite{MR1908783}).

The second open question concerns the expressive power of the quantifiers $Q^{\star}_L$, for a  regular $L$. It is an open question whether non-regular languages can be defined in   $\SOM(\QmodM)$.

\bibliographystyle{abbrv}
\bibliography{Leafbib}

\end{document}